\documentclass[prb,twocolumn,showpacs,amsmath,amssymb]{revtex4-1}
\usepackage{graphicx} 
\usepackage{epsfig}
\usepackage{dcolumn} 
\usepackage{color}
\newcommand{\beq}{\begin{equation}}
\newcommand{\eeq}{\end{equation}}
\newcommand{\beqa}{\begin{eqnarray}}
\newcommand{\eeqa}{\end{eqnarray}}

\begin{document}
\title{Negative electronic compressibility and nanoscale inhomogeneity in 
ionic-liquid gated two-dimensional superconductors}
\author{G. Dezi$^1$, N. Scopigno$^{1,3}$, S. Caprara$^{1,2}$, and M. Grilli$^{1,2,*}$}
\affiliation{$^1$Dipartimento di Fisica Universit\`a di Roma Sapienza P.$^{le}$ Aldo Moro 5 00185 Roma, Italy\\
$^2$ISC-CNR  Unit\`a di Roma Sapienza\\
$^3$ Institute for Theoretical Physics  3584 CC Utrecht the Netherlands \\
$^*$ Correspondence and
requests for materials should be addressed to M.G.(marco.grilli@roma1.infn.it)}

\begin{abstract}
When the electron density of highly crystalline thin films is tuned by chemical doping or ionic liquid 
gating, interesting effects appear including unconventional superconductivity, sizeable spin-orbit 
coupling, competition with charge-density waves, and a debated low-temperature metallic state that seems 
to avoid the superconducting or insulating fate of standard two-dimensional electron systems. Some experiments 
also find a marked tendency to  a negative electronic compressibility. We suggest that this indicates an 
inclination for  electronic phase separation resulting in a nanoscopic inhomogeneity. Although the mild 
modulation of the inhomogeneous landscape is compatible with a high electron mobility in the metallic state, 
this intrinsically inhomogeneous character is highlighted by the peculiar behaviour of the 
metal-to-superconductor transition. Modelling the system with superconducting puddles embedded in a  
metallic matrix, we fit the peculiar resistance vs. temperature curves of systems like TiSe$_2$, MoS$_2$, and 
ZrNCl. In this framework also the low-temperature debated metallic state finds a natural explanation in terms 
of the pristine metallic background embedding non-percolating superconducting clusters. An intrinsically 
inhomogeneous character naturally raises the question of the formation mechanism(s). We propose a mechanism 
based on the interplay between electrons and the charges of the gating ionic liquid.

\end{abstract}

\maketitle

\vspace{0.5 truecm}
\section{introduction}
\begin{figure*}[htbp]
\includegraphics[angle=-0,scale=0.6]{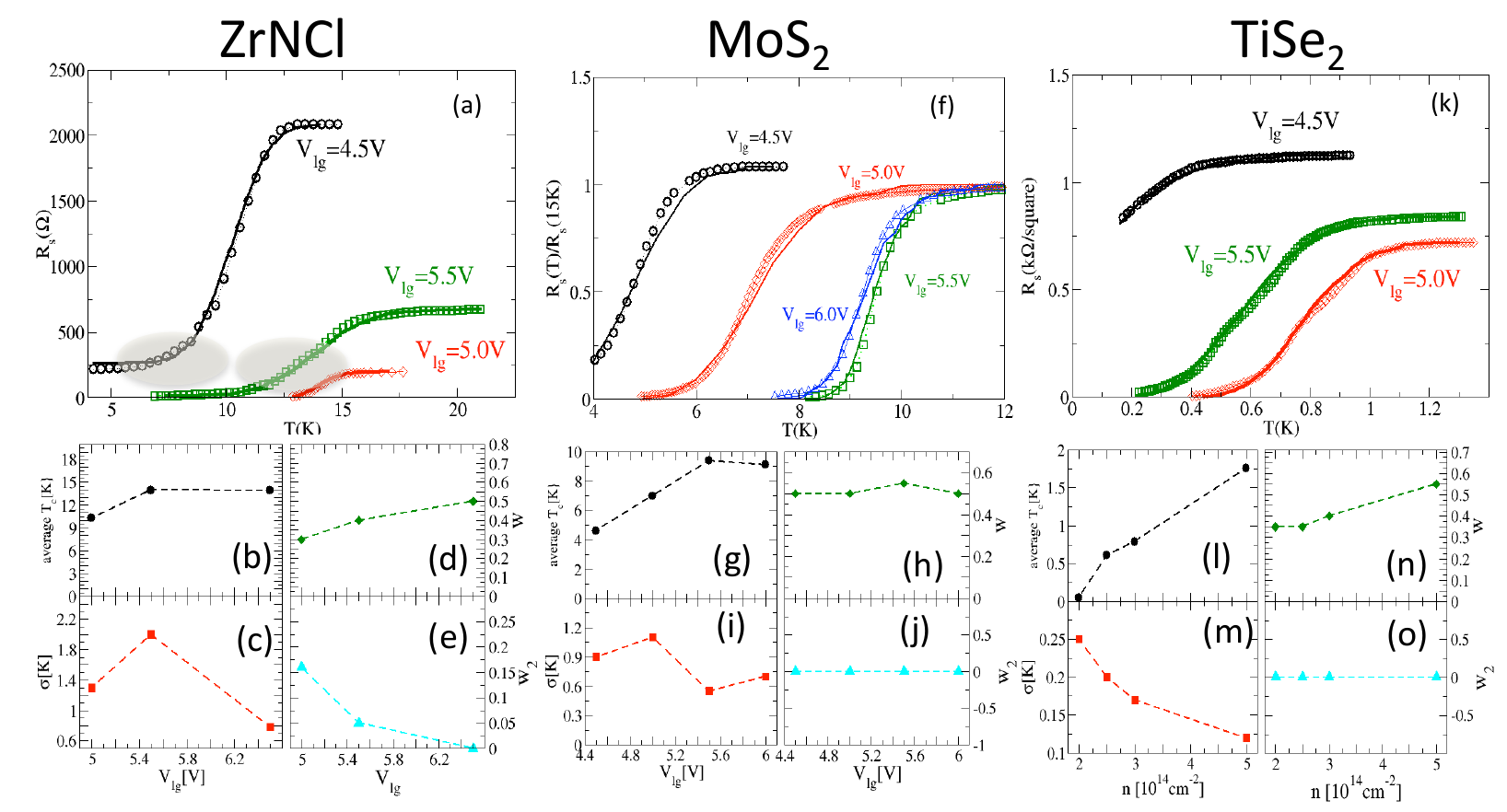}
\caption{Fit (symbols) and experimental resistance (solid curves) (a) of ZrNCl (from 
Ref.\,\onlinecite{Saito-2015}) at three different values of the ionic-liquid gating ($V_{IG}=$5, 5.5, and 
6.5\,V).  (b-e) The average local ${\overline{T}_c}$, the variance $\sigma$ of their distribution, the total 
weight $w$ of the superconducting regions and the fraction $w_2$ of broken bonds in the filamentary 
superconducting cluster (see Appendix A) as resulting from the fits. (f-j) Same as in (a-e) for the MoS$_2$ 
experiments of Ref.\,\onlinecite{Ye-2012}. (k-o) Same as (a-e) and (f-j) for the TiSe$_2$ experiments of 
Ref.\,\onlinecite{Li-2015}. The shaded regions in (a) highlight the tailish character of the resistance 
curves and the saturating plateau at low electron density and temperature, marking the regime without a 
percolating superconducting subset.
}
\label{resist}
\end{figure*}

In last years, great advances have been achieved in the fabrication of idealized two-dimensional (2D) 
electron systems such as heterogeneous interfaces, molecular-beam-epitaxy grown atomic layers, exfoliated 
thin flakes and field-effect devices. Moreover, the possibility of combining or singling out isolated layers 
of graphene, transition metal dichalcogenides, high-temperature superconducting cuprates, transition metal 
oxide interfaces, and so on, has opened new perspectives of a `Lego'  functionalization of 2D systems (see, 
e.g., Ref.\,\onlinecite{geim2013}). One common relevant feature of these systems is their highly crystalline 
character, which renders them an ideal playground for studying several intriguing physical 
effects,\cite{iwasa-2016} like, e.g., Ising superconductivity or topological phases,
without the complications due to impurity or defect-induced disorder. Among these properties there is the 
appearance of a low-temperature quantum metallic state, which has been proposed as a new state of matter
escaping the standard dichotomy of 2D electron systems, which at low temperature are usually forced to 
choose between an insulating (typically due to Anderson localization) or a superconducting state. These 
highly crystalline 2D systems, like transition metal dichalcogenides or ZrNCl ultra-thin films, instead, 
display a low-temperature resistance saturation. They start from a metallic high-temperature state, then 
display a suppression of resistance due to the occurrence of superconductivity, when the temperature $T$ 
is lowered. However, when superconductivity is (partially) destroyed by a magnetic field, they display a 
low-$T$ metallic behaviour, often marked by a rather extended plateau, with the resistance extrapolating 
to a finite constant value. There are some features of the resistance curves $R(T)$ that are quite specific 
of this anomalous state. First of all, the transition from high-$T$ metal to superconductor (or to low-$T$ 
metal) is very broad, so broad that by no means its width can be reproduced by standard superconducting 
fluctuations mechanisms {\sl \`a la} Aslamazov-Larkin or Halperin-Nelson.
Second, the shape of $R(T)$ is peculiarly tailish with a long foot in its low-$T$ side. Traditionally 
this behaviour is attributed to vortex dissipation,\cite{tinkham} but, remarkably, the very same features 
occur in the absence of magnetic field, when superconductivity is weakened or partially suppressed by 
reducing the electron density. This indicates that the usual mechanism of vortex dissipation cannot be 
invoked. The case of ZrNCl reported in Fig.\,\ref{resist} (a), where these features 
[$R(T)$ saturation and tailish behavior] are highlighted by a grey shadowing, is particularly eloquent.
 
A possible clue to interpret these unusual features comes from the comparison with the 2D electron gas 
formed at the oxide interfaces like, e.g., LaAlO$_3$/SrTiO$_3$, where the temperature dependence of the 
resistance curves display a very similar behaviour. Despite their highly crystalline structure, the above 
peculiar features have been successfully explained in these systems in terms of nanoscale electronic 
inhomogeneities.\cite{biscaras-2013,bucheli-2013,caprara-2013,guenevere-2016}
Further indications that oxide interfaces and other 2D electron systems are peculiarly inhomogeneous come 
from the observation of a quantum Griffiths state when a perpendicular magnetic field drives a 
metal-to-superconductor transition.\cite{shengchun-2016,saito-2018}
Of course, inhomogeneity might seem at odds with the remarkably ordered structure of these systems and 
the high mobility of the charge carriers indicates that the usual low-$T$ scattering mechanisms (crystal 
defects and impurities) are not of primary relevance in these systems. This suggests that inhomogeneities are 
not induced by extrinsic sources, but likely occur from intrinsic mechanisms\cite{Caprara-2012,Scopigno-2016} 
that destabilise the electronic liquid giving rise to (short-scale) electronic phase separation and density 
fluctuations on the nanoscopic scale. 

In this work we consider transport experiments and, from the analysis of resistivity curves near the 
metal-to-superconductor transition, we provide a clear evidence of the inhomogeneous character of some 2D 
highly crystalline systems like transition metal dichalcogenides and ZrNCl, and we deduce the structure of 
this inhomogeneity. We focus on the metal-to-superconductor transition driven by changing the electron density 
in the absence of magnetic field, so that the observed transport properties should be interpreted without 
invoking the dissipation effects of vortices. The model we adopt, although based on reasonable assumptions, 
is phenomenological in nature and does not aim at identifying the microscopic mechanisms underlying the 
inhomogeneity formation, its structural and electronic properties and so on. It rather aims at introducing 
the minimal amount of knobs to be tuned in order to reproduce the data and to extract information on the
inhomogeneity (the fraction of high-vs-low density regions, the connectivity of the inhomogeneous clusters, 
the way superconductivity disappears when the average density is reduced, and so on).

In the second part of this paper, we address the origin of the inhomogeneity and the possibility that it 
arises from an electronic phase separation, as indicated by experimental evidences of negative electronic 
compressibility.\cite{tutuc-2014,King-2015} In this framework, we also propose a microscopic mechanism of 
electronic instability based on the interplay between the electron gas and the ionic countercharges due to 
ionic liquid gating. 

\section{The inhomogeneous transport model}
\subsection{The physical scenario of inhomogeneous 2D crystalline superconductors}
Our work moves from two phenomenological observations: a) the metal-to-superconductor transition is 
generically so broad in 2D crystalline superconductors that no sensible fluctuation mechanism can account for 
it\cite{Ye-2012,Saito-2015,Li-2015} [see Fig.\,\ref{resist} (a,f,k)]; b) when the filling of the 2D electron gas 
is induced by ionic liquid gating, the width of the transition is generically broader and it is always 
accompanied by a tailish character in the low-$T$ part of the resistance-vs-temperature curves $R(T)$ [see, 
e.g., the shaded regions in Fig.\,\ref{resist}(a)]. This latter feature is much less pronounced in chemically 
doped systems.\cite{Bhoi-2017}

To account for these observations, we assume that the electron gas in 2D crystalline superconductors is 
inhomogeneous, with a metallic matrix hosting puddles that become superconducting below a random local 
critical temperature $T_c$.  
\begin{figure}[htbp]
\includegraphics[angle=-0,scale=0.25]{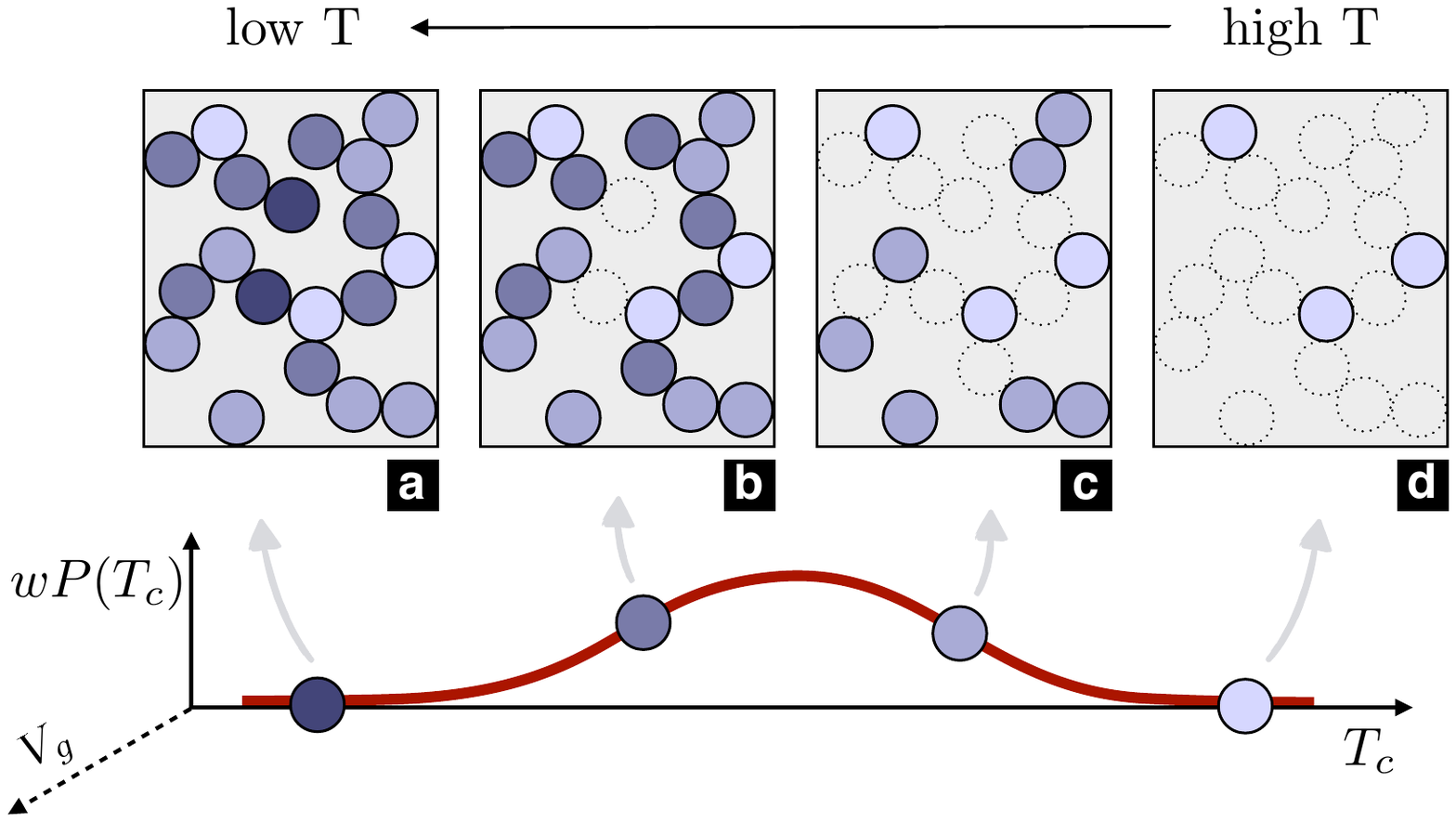}
\caption{ (Top) Schematic representation of an inhomogeneous 2D crystalline superconductor. At high temperature 
the metal is weakly inhomogeneous. As the temperature is reduced, from right to left, more and more 
superconducting puddles (the light-to-dark blue circles) are found with a critical temperature $T_c>T$, as 
highlighted by increasingly darker hues of color. Correspondingly, an increasing fraction of the system
becomes superconducting. A zero resistance state is reached as soon as a percolating path is formed. (Bottom) 
The number of superconducting puddles is  ruled by the probability distribution $P(T_c)$ of the critical 
temperatures and $w$ is the overall fraction of potentially superconducting regions. The distribution and 
fraction may depend on control parameters like, e.g., the gate voltage $V_g$ tuning the electron density of the system. 
}
\label{scheme1}
\end{figure}
The length scale of the inhomogeneity is immaterial, provided the puddles are 
large enough to sustain a superconducting state below the given local $T_c$ and the metal-superconductor 
mixture is fine enough to allow for a good statistical sampling even on small samples a few micrometers large 
(in oxide interfaces like, e.g., LaAlO$_3$/SrTiO$_3$, such conditions are met with typical inhomogeneities on 
the scale of a few hundreds of nanometers,\cite{biscaras-2013,bucheli-2013,caprara-2013,guenevere-2016} 
slightly larger than the superconducting coherence length $\xi\approx 50$\,nm). 
Before presenting any specific model, we now find it useful to give a pictorial view of the physical 
situation.

{\it {---The inhomogeneous metallic state ---}}
 First of all, we {\it assume} that the system has an inhomogeneous density. This density 
inhomogeneity has minor consequences in the metallic phase, because quasiparticles are weakly scattered by 
extended inhomogeneities. Although a quantum theory for transport in inhomogeneous media is still lacking, 
the effects of scattering due to extended impurities has already been considered in the past (see, e.g., 
Ref. \onlinecite{zhu2004}). It is known that when the extension of the inhomogeneity is large enough, 
it hardly affects the mobility of carriers because electrons, when crossing large inhomogeneities at 
(slightly) different density, are not backscattered (which would strongly degrade currents), rather they are
weakly refracted, giving rise to a dominantly forward scattering, which is not detrimental for transport. In 
other words, metallic states with weakly modulated densities can still display high charge mobility and 
their transport properties are hardly different from those of clean homogeneous metals. In a 
phenomenological classical scheme like, e.g., the effective medium theory,\cite{landauer,kirkpatrick} a mixture 
of metals with slightly different local resistivities cannot be distinguished from a homogeneous metal with 
the resulting effective resistivity.
Owing to the extended nature of disorder in these systems, we also expect that localization effects (usually coming from
interference effects in time-reversed electronic path scattered by quenched point-like impurities) have minor relevance.

{\it {---The inhomogeneous superconducting state ---}}
The above situation becomes drastically different when some regions start to become superconducting by 
lowering $T$. The contrast between higher- and lower-density regions is then strongly enhanced because some 
regions abruptly acquire a vanishing resistance. In this situation superconducting puddles at higher 
electron density are embedded in a metallic matrix with large effect on the overall resistivity and other 
transport properties. In this regard, we notice that several mechanisms can trigger a local superconductivity 
with even a small local increase of the electron density. In LaAlO$_3$/SrTiO$_3$ interfaces, for instance, it 
is known that a superconducting dome arises in the 2D electron gas when the filling is increased and the
Fermi energy exceeds a threshold $E_c$ allowing the occupation of additional bands with higher density of 
states.\cite{biscaras-2012,joshua-2012} An alternative mechanism could arise from the interplay with 
a competing phase as it often occurs near a quantum phase transition, where superconductivity, being favoured 
by quantum fluctuations and by the weakening of the competing order, forms a superconducting dome in the 
phase diagram region around the quantum critical point. These arguments should justify our assumption that
 even a rather weak inhomogeneity in electron density may result  at 
low temperature into an inhomogeneous mixture of normal and superconducting regions. 

{\it {--- The distribution of local critical temperatures ---}}
We now discuss the corresponding distribution of local critical temperatures. In the context of 
LaAlO$_3$/SrTiO$_3$ interface, a simple calculation within BCS theory\cite{caprara-2013} shows that a rapid 
increase of the local T$_c$ takes place when the chemical potential passes the energy threshold  for filling 
the bands relevant for superconductivity, $T_c\propto \sqrt {\mu -E_c}$. Therefore, weak density
variations induce sizeable variations on $T_c$. In this framework one can also generically notice that the 
global $T_c$ can vary rapidly with small variations of the average electron density (see, e.g., Fig. 7 of 
Ref.\,\onlinecite{iwasa-2016}, where the phase diagrams of ZrNCl and MoS$_2$ are reported). 
These arguments suggest that mild spatial variation of the density may induce a rather broad distribution of local
$T_ c$s.

Although disorder is not a main actor in highly crystalline films, it can still have some effects on the 
local $T_c$s: It is known that $T_c$ can be substantially reduced by disorder\cite{finkelstein-1987} and it 
is conceivable that fluctuations in the spatial distribution of impurities and/or of the local chemical 
potential (due to the density inhomogeneity) at the nanoscale induce  variations of $T_c$ in the superconducting puddles.
This provides an additional argument to support the idea that {\it the superconducting puddles may have a rather large 
distribution $P(T_c)$ of local superconducting critical temperatures}. As schematically depicted in 
Fig.\,\ref{scheme1}, the resistance of the system keeps varying (and possibly vanishes if a percolating 
cluster of superconducting puddles is formed) when $T$ is progressively decreased around the 
metal-to-superconductor transition. The width of the $T_c$ distribution is then directly related to the width 
$\sigma$ of the average metal-to-superconductor (or high-$T$--to--low-$T$ metal) transition. 

In summary, the scenario we consider is characterised by: a) a weak density inhomogeneity that is compatible 
with a high-mobility metallic state; b) a mixed metal-superconductor state with a broad transition ruled by 
the width of the $T_c$ distribution of the superconducting puddles; c) the possible occurrence of 
low-$T$ metallic state as a mixture of normal and superconducting regions in the absence of a percolating 
superconducting cluster: as soon as all the potentially superconducting regions have zero resistance, the 
overall resistance no longer varies by further decreasing $T$. Due to the highly crystalline character 
other mechanisms, like, e.g., Anderson localisation, are still not effective at the lowest experimentally accessed
temperatures and the resistance appears to 
saturate to a constant value. In this way the mysterious low-$T$ quantum metal state acquires a quite natural 
and simple interpretation: it is just the metallic state of a very clean system in the presence of embedded, 
non-percolating superconducting puddles. Other specific features, like, e.g., the tailish $R(T)$, can also 
be reproduced, as they result from specific structural features of the inhomogeneous clusters.
The study of these features of course requires a specific model, that we introduce in the next subsection.

\subsection{The Random-Resistor Network model}
To describe the above physical scenario, we represent the system by a random-resistor network (RRN). 
We have to capture somewhat opposite characteristics of the data. On the one hand, the resistance curves 
tend to vanish (or to saturate at a finite value, if percolation does not occur) with a rather long tail, 
which is the hallmark of a {\it weak long-distance connectivity of the superconducting 
cluster}.\cite{bucheli-2013,caprara-2013,CGBC} This is because in a (nearly) one-dimensional conductor, a zero-resistance
superconducting state is only achieved when {\it all} the bonds have become superconducting. This requires a more stringent condition that
is realised at a lower temperature than the average $T_c$ of the random critical temperature distribution: until the very last
resistive bonds are switched off R stays finite giving rise to the low, tailish shape of $R(T)$. 

\begin{figure*}[htbp]
\includegraphics[angle=-0,scale=0.60]{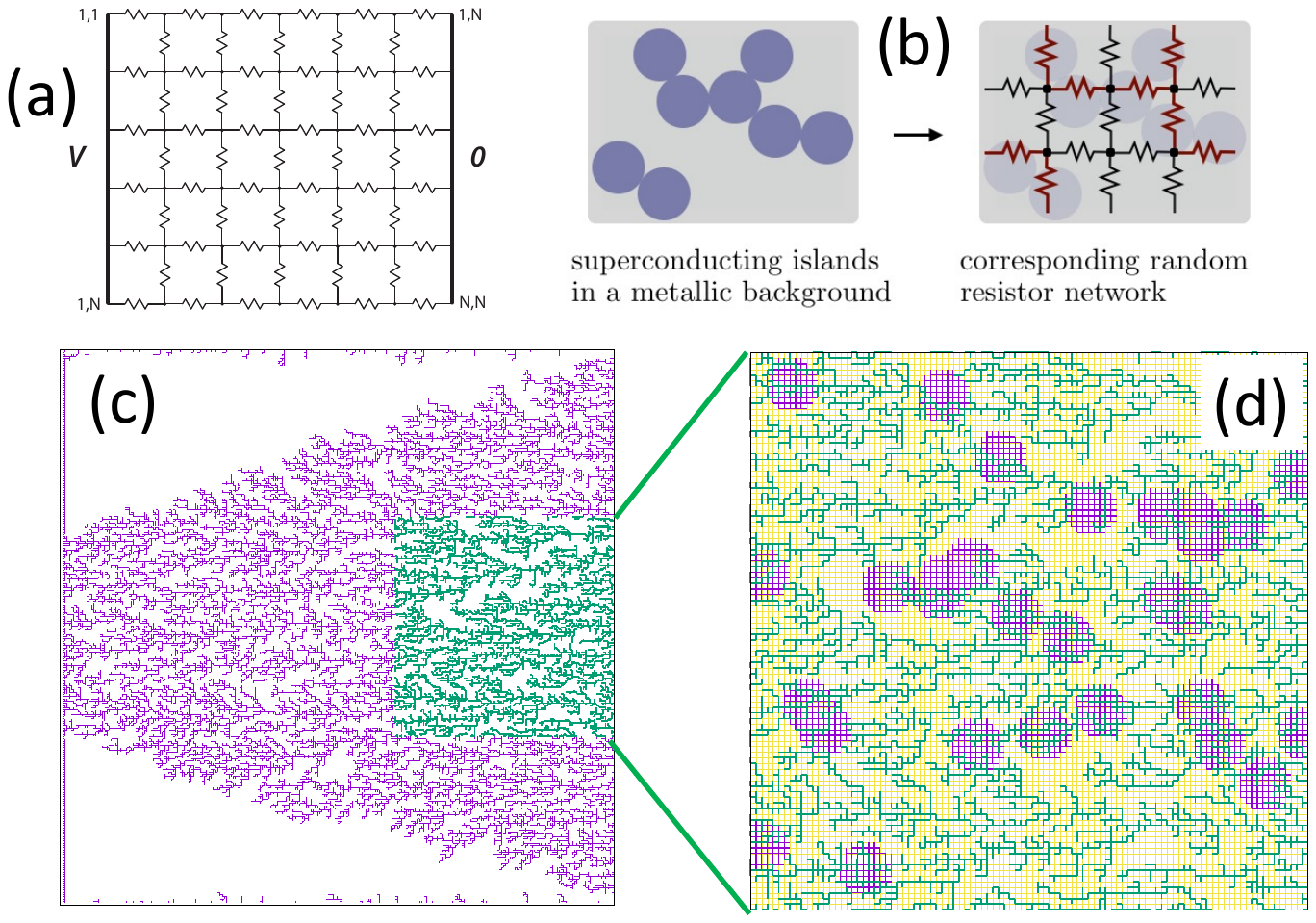}
\caption{ (Color online) (a) Schematic view of a $N\times N$ resistor network. The left and 
right vertical edges are kept at potential $V$ and $0$, respectively.
(b) Scheme of the mapping of a system of superconducting islands embedded in a metallic
background onto a random resistor network. The right (b) panel is an enlargement of the 2D square lattice of
metallic bonds: The resistors highlighted in red are those that become superconducting below
a random local critical temperature $T_c$. (c) Example of a filamentary structure produced by letting 50000 
particles diffuse across a $250 \times 250$ square lattice, according to the DLA prescription explained in 
Appendix A. Notice that the underlying 2D square-lattice grid is not reported for clarity (this is the overall 
blank part of the figure) and only its DLA fractal subset is represented. From this larger fractal a restricted 
$100 \times 100$ square sub-lattice is extracted from the original $250 \times 250$, to serve as a 
filamentary skeleton for the superconducting component of our RRN. (d) Example of a superconducting cluster 
of our RRN, obtained superimposing bulkier circular superpuddles (in purple color), of diameter equal to 10 bonds, to the 
(green) fractal skeleton generated by means of the DLA prescription. In this panel the underlying 2D square-lattice grid is
reproduced in light yellow.}
\label{scheme1B}
\end{figure*}

On the other hand, when lowering the temperature, 
already well above the transition, there is a marked decrease of the resistance, which extends over a 
broad temperature range. This indicates that {\it a whole substantial part of the system is becoming superconducting}. 
To capture this multi-faceted physics, we should therefore account for three distinct 
features, namely the weak long-distance connectivity of the superconducting network, its bulky character, 
and the randomness of the critical temperature of its constituents. To this purpose, we conceived the 
following composite spatially correlated structure (for details see Appendix A): 

1) First of all we discretise the 2D inhomogeneous metallic gas by taking a 2D square lattice where each 
resistor is located on the $i$-th bond of the grid [Fig.\,\ref{scheme1B} (a)] and can either stay metallic with a typical resistance 
$R_0$ down to the lowest accessible temperature, or become superconducting (i.e., its resistance vanishes) 
below a given local critical temperature $T_{c}(i)$, randomly extracted from a probability distribution.  
Fig.\,\ref{scheme1B} (b, right panel) displays an enlarged view of the whole 2D square lattice of 
resistors. The black ones represent the metallic matrix and keep their resistance $R_0$ finite down to the 
lowest temperature, while the red resistors belong to the potentially superconducting clusters and become superconducting below their local $T_c(i)$. 

It turns out that the specific shape of the $R(T)$ curves can only be described by a {\it spatially correlated} 
cluster,\cite{CGBC,bucheli-2013} and luckily its features are quite informative about this spatial structure.
 
2) To distribute spatially these potentially superconducting bonds [the red resistors of Fig.\,\ref{scheme1B}
(b, right panel)], we generate a fractal-like structure using a Diffusion Limited Aggregation (DLA) algorithm 
and we select a region of the fractal cluster filling our numerical cluster (the green subset in Fig. \ref{scheme1B} (c). 
This fractal character by no means 
implies that real systems have a self-similar spatial distribution of superconducting regions, but it is a 
mere technical tool to generate a spatially correlated cluster with a random and filamentary structure. On top 
of this faint filamentary skeleton, circular (i.e., bulkier) large puddles (we nickname them {\em superpuddles}) 
are randomly added [the purple circular regions in Fig. \ref{scheme1B}(d)]: 
We randomly select a superconducting bond of the skeleton and we manually add more 
superconducting bonds in the vicinity to form a circular patch of superconducting bonds. All the bonds (i.e., 
the resistors) in the superpuddle are set to have the same random local $T_c$. The addition of superpuddles stops when a total 
weight $w$ is reached to attain a given superconducting fraction in the system (this is one of our fitting 
parameters). Figs.\,\ref{scheme1B}(d)  and \ref{device} (c) display typical structures of the RRN inhomogeneous 
cluster, where the filamentary skeleton coexists with the circular superpuddles. 

When the superconducting cluster does not percolate the resistance curve $R(T)$ at low $T$ saturates at a finite value.
This occurs because the total superconducting weight in the system is low and some breaks occur in  the fractal filaments
spoiling the connectivity of the superconducting cluster.  To describe this situation, 
we randomly chose some resistors in the filamentary fractal subset of the 2D square-lattice grid and turn 
them back to metallic (i.e., we manually set their local $T_{c}(i)=0$). The fraction $w_2$ of these broken 
bonds is also adjusted by the fit.

3) Once the above complex structure of the superconducting cluster is generated, the model is completed by 
assigning a probability distribution of the random local critical temperatures. For the sake of definiteness 
we adopt a Gaussian distribution
\begin{equation}
P(T_c) =\frac{1}{\sqrt{2\pi}\sigma}\,\mathrm e^{-\frac{\left(T_c-\overline{T}_c\right)^2}{2\sigma^2}}.
\label{gauss}
\end{equation}

\begin{figure}[htbp]
\includegraphics[angle=-0,scale=0.30]{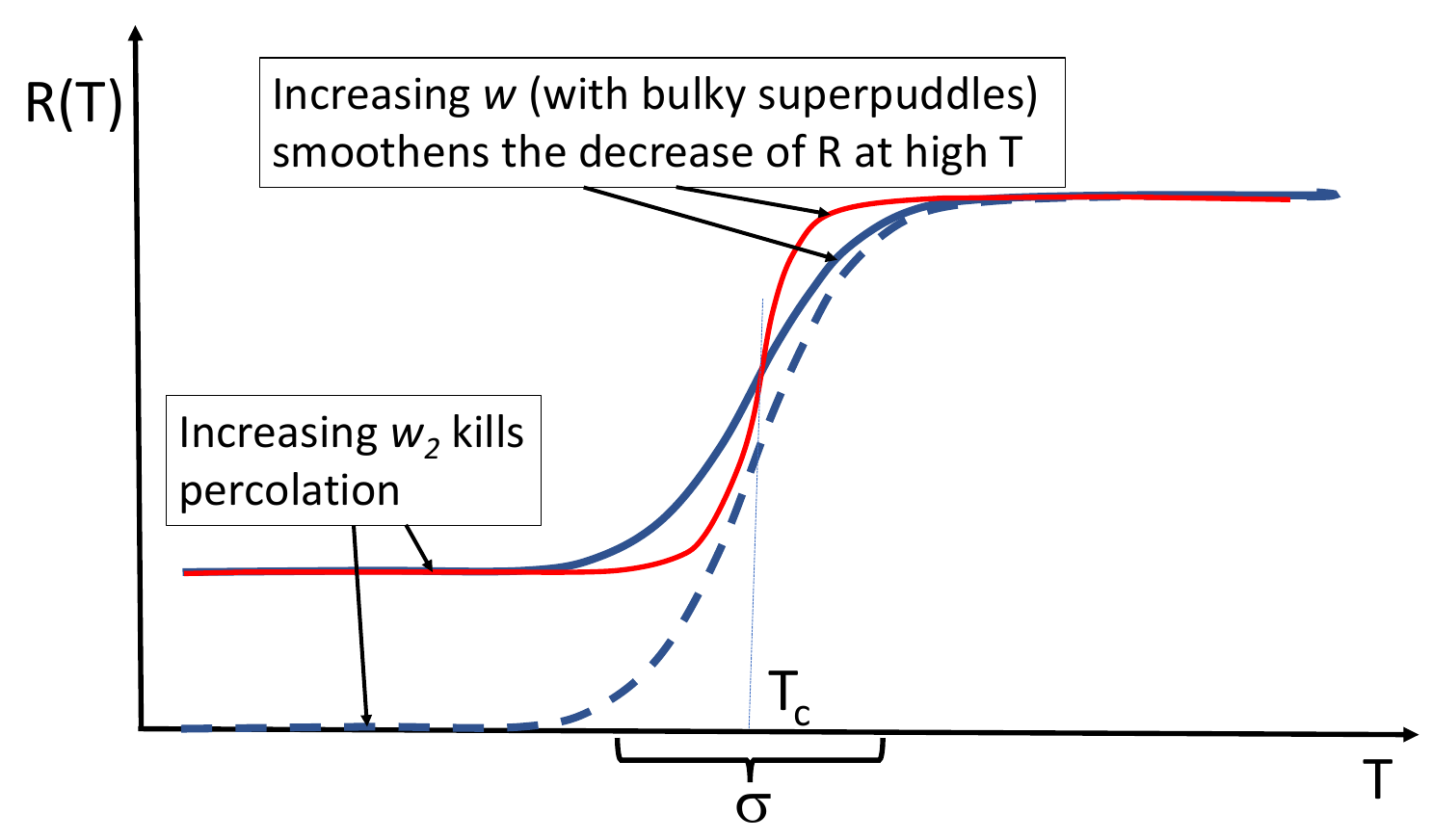}
\caption{ (Color online) Schematic view of how the shape of $R(T)$ depends on the various parameters of the 
model. Increasing $w_2$ while keeping $w$ fixed, more superconducting filaments are broken and percolation is 
lost: The dashed blue line becomes the solid blue line. Increasing $w$ by increasing the number of 
superpuddles renders more gradual the decrease of $R(T)$ (red-to-blue solid lines). $\overline{T}_c$ is the 
center of the transition, while $\sigma$ rules its width.
}
\label{scheme1C}
\end{figure}

In summary, our RRN model is characterised by four parameters ($\overline{T}_c$, $\sigma$, $w$, $w_2$) 
each having its specific effect on the resistance curve and therefore providing information on the structure 
of the superconducting clusters. Specifically, as schematically described in Fig.\,\ref{scheme1C}, 
$\overline{T}_c$ and $\sigma$ rule the position and the width of the resistance decrease around the 
metal-to-superconductor (or high-$T$-to-low-$T$ metal) transition. $w$ is the fraction of superconducting 
regions and, being mostly arranged by the number and size of the superpuddles, it determines the smoothness 
and gradualness of the decrease of $R(T)$ in the high-$T$ side of the curve. We also notice that the radius of 
the superpuddles is not a relevant parameter as long as it is not too large in comparison with the size of the 
RRN (if this not the case, only a few superpuddles are contained in the cluster and steps appear
in $R(T)$ whenever a single  superpuddle becomes superconducting). Finally, $w_2$ is the most effective parameter 
in ruling the long-distance connectivity of the superconducting clusters. 

We emphasise that both the bulky superpuddles and the filamentary connections are crucial to reproduce the shape 
of $R(T)$: the former give a smooth substantial decrease starting at high temperature, while the latter are 
responsible for the tailish shape at low temperatures.

\subsection{Using the RRN to fit the resistance data of 2D superconductors}
Once the random structure of the inhomogeneous RRN is set, we then devote a systematic numerical effort 
to solve the Ohm's and Kirchhoff's equations on bonds and nodes of the clusters (of typical size $200\times 200$,
see Appendix A), thereby determining the local and global currents and voltages, hence
the global resistance of the system, to fit the experimental curves. While the results of the fits are given in 
Fig.\,\ref{resist} (a),(f), and (k) for ZrNCl, MoS$_2$, and TiSe$_2$ respectively, the corresponding 
values of the parameters characterising the cluster geometry ($w$ and $w_2$) and the $T_{c}(i)$ 
distribution ($\overline{T}_c$ and $\sigma$) are reported in the lower panels of Fig.\,\ref{resist}.

Clearly, our model captures the tailish character of the $R(T)$ curves near the zero-resistance limit, as well 
as the saturation to finite values when percolation is not achieved. Both these effect are a consequence of the 
poor long-distance connectivity of our superconducting cluster: the system has no chance to become superconducting 
even when all resistors inside the subset are switched off and the resistance at low $T$ does not vanish 
and saturates at a finite value because of just a few residual metallic bonds. Therefore, within this 
percolative scheme, the residual finite resistance remaining at low temperature in some low-density samples has 
a very natural interpretation: it is due to the pristine metallic matrix embedding the (non-percolating) superconducting puddles. 

\begin{figure*}[htbp]
\includegraphics[angle=-0,scale=0.55]{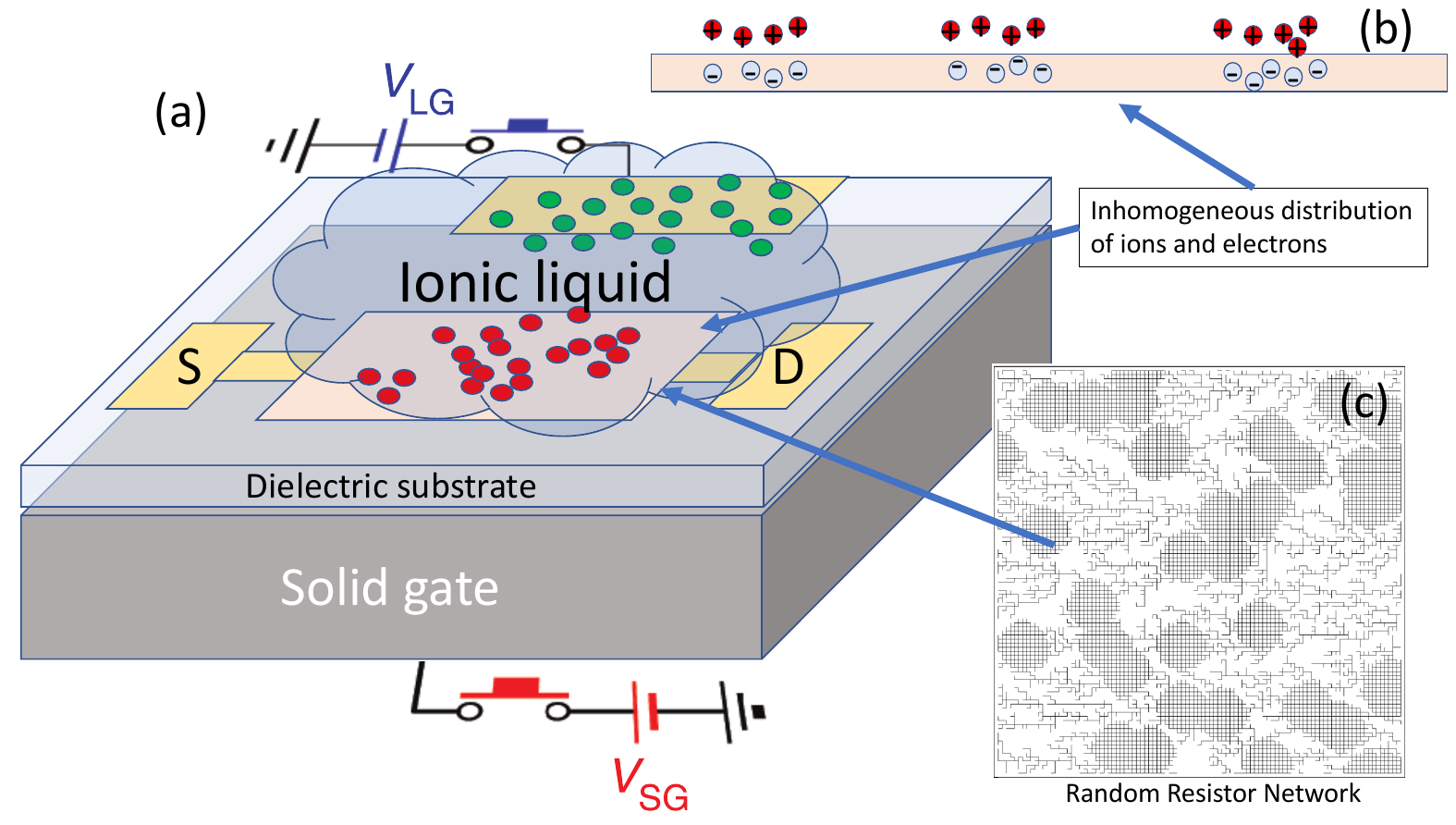}
\caption{(a) Schematic representation of a 2D crystalline superconductors (pink rectangle) on a dielectric 
substrate in the presence of a metallic back gate (grey region) and a ionic liquid (light-blue droplet). 
The positive (red) and negative (green) ions are also represented together with the ionic liquid gating 
(light orange rectangle). S and D are the source and drain electrodes. (b) Schematic profile of the 
inhomogeneously doped layer of a 2D crystalline superconductor. (c) $100 \times 100$ cluster of RRN showing 
the filamentary structure together with the bulkier superpuddles.}
\label{device}
\end{figure*}

The model also allows to distinguish two physically different situations: in one case superconductivity 
disappears at low $T$ because upon reducing the average electron density the superconducting puddles 
become less dense and more sparse, while in the second case the disappearance of superconductivity is 
driven predominantly by a reduction of the average superconducting critical temperature in the puddles, 
e.g., because of some competing mechanism. Both these situations are found to occur and are reported in 
Fig.\,\ref{resist}. In particular, one can see that in ZrNCl [Fig.\,\ref{resist} (a)] the average local 
critical temperatures only varies from 14\,K at high density to 10\,K at the lowest density 
[Fig.\,\ref{resist}(b)]. Notice that in this case the superconducting fraction [Fig.\,\ref{resist}(d)] is 
so low that the puddles do not percolate and the system stays metallic down to the lowest temperatures. 
Therefore, the system fails to reach the zero-resistance state even though a substantial part of it is locally
superconducting with a rather large $T_{c}\approx 10$\,K [Fig.\,\ref{resist}(b)]. The moderate reduction of 
the average $T_c$ [Fig.\,\ref{resist}(b)] can be explained because the global average reduction of electron 
density reflects in a reduction of the local density even in the superconducting puddles, thereby inducing 
a relatively larger effect of quenched impurities. The RRN model even allows to distinguish whether the 
density reduction affects more the superpuddles (see Figs.\,\ref{scheme1B}(d)  and \ref{device} (c)) or 
the connecting superconducting filaments. One can see that the high-$T$ parts of the resistance curves at 
the transition are quite similar, indicating that the bulky part of the superconducting regions is more or 
less unaffected, while the density decrease affects more the connecting filaments, which disappear with a 
fraction $w_2$, which increases while reducing the gating [Fig.\,\ref{resist}(e)]. Of course, some more 
rapid downward bending of the black curve at $V_{lg}=4.5$\,V than in the $V_{lg}=5.0,\,5.5$\,V cases, 
indicates that the density decrease also reduces the weight of the superpuddles, but this is comparatively 
less relevant for transport. In other words we see that for ZrNCl percolation is destroyed because the 
underlying geometrical support is strongly affected (namely many connecting bonds no longer superconduct, 
see the increase of $w_2$ by reducing the overall density), while the distribution of local $T_{c}$s 
is modified in a comparatively less relevant manner.

On the other hand, transition metal dichalcogenides are notoriously characterized by a strong tendency to 
form CDWs, which compete with superconductivity. Although it is overwhelmingly difficult to microscopically 
describe this interplay and competition, it is clear that reducing the average density in these systems 
strengthens the CDW order, which has a strong influence also inside the superconducting puddles. Therefore, 
the local $T_{c}$s are suppressed and vanish at low enough electron density. In this case, the disappearance 
of the superconducting phase does not occur because of a lack of percolation, but because the weakest part of 
the superconducting subset loses its superconducting character. This can be easily recognised in the lack of 
a saturating low-$T$ resistance in Figs.\,\ref{resist}(f,k) for MoS$_2$ and TiSe$_2$. In this case, 
therefore, contrary to the ZrNCl case, the geometrical support of superconductivity is less affected, while 
the local $T_{c}$s are generically reduced by the competition with CDW and their overall distribution is 
shifted to lower values. 

\section{ The mechanisms of electronic phase separation}
Once the inhomogeneous character of the ionic-liquid gated 2D crystalline superconductors is assessed 
via the above phenomenological analysis, the crucial question is left about the origin of this inhomogeneity. 
Due to the quite general occurrence of such inhomogeneity in different systems doped by ionic liquid gating, 
we propose here that the 2D electron gas in these systems may become thermodynamically unstable and undergo 
an electronic phase separation, thereby displaying a negative compressibility due to the combined action 
of the confining potential well and the gating ionic countercharges. Of course, as it will be discussed 
in the last part of this paper, several mechanisms intervene to stop the full development of the electronic 
phase separation, giving rise to a final thermodynamically stable (but inhomogeneous) system. Here we only 
focus on the electronic part of the system and its interplay with the ions of the liquid gate to explain 
the source of instability. We describe the 2D electron gas as a free electron gas confined in a potential 
well that quantises the electron motion in the direction perpendicular to the interface. The depth of the 
well depends on the amount of countercharges (i.e., the positive ions in the liquid nearby the 2D 
crystalline superconductor) per 2D unit cell, $\nu$, so that the band dispersion of the 2D electron gas will 
be henceforth written as $\varepsilon_k=\varepsilon_0(\nu)+\frac{k^2}{2m}$, where 
$\varepsilon_0(\nu)$ is the quantized level in the confining potential well, above which the 2D band 
dispersion arises, $m$ is a suitable effective mass, and $k$ is the 2D quasimomentum parallel to the interface. 
For a given density of countercharges $\nu$, the grand-canonical electronic Hamiltonian reads
\beq
{\cal H}=\sum_{k,\sigma}\left(  \varepsilon_k-\mu    \right) \hat n_{k,\sigma} -\lambda \left(  \sum_{k,\sigma}  \hat n_{k,\sigma}-N\nu    \right),
\eeq
where $\mu$ is the electron chemical potential, $N$ is the number of cells,  and the Lagrange multiplier $\lambda$ 
enforces the constraint  that each electron comes from a dopant countercharge, i.e. the average electron density $n=\nu$. 
${\hat n}_{k,\sigma}$ is the operator that counts the number of electrons of quasimomentum $k$ and spin $\sigma$. 

We assign a bandwidth $W$ to the 2D electron gas, and write the DOS as $N_0=1/W$. Then the condition of 
overall neutrality at $T=0$ becomes
\beq
n=\frac{2}{W}\int_{\varepsilon_0(\nu)}^{\mu+\lambda}\,\mathrm d\varepsilon=
\nu 
\label{constraint}
\eeq
since the Fermi statistics counts the states with negative eigenvalues of ${\cal H}$, $\varepsilon_k-\mu-\lambda<0$,
the minimum value of $\varepsilon_k$ being $\varepsilon_0(\nu)$. The factor of 2 counts for the spin multiplicity.
Then,  the $T=0$ grand-canonical thermodynamical potential per unit cell reads
\begin{eqnarray}
\omega&=&\frac{2}{W}\int_{\varepsilon_0(\nu)}^{\mu+\lambda}\left(\varepsilon-\mu-\lambda\right)\,
\mathrm d\varepsilon+\lambda \nu+\omega_0(\nu)\nonumber\\
&=&-\frac{1}{W}\left[\mu+\lambda-\varepsilon_0(\nu)\right]^2+\lambda \nu+\omega_0(\nu), 
\label{grandcanonical}
\end{eqnarray}
where $\omega_0(\nu)=\frac{1}{2}A\nu^2+\frac{1}{4}B \nu^4$ is the countercharge contribution, the first 
term resulting from the countercharge inverse compressibility, and the second term modeling the cost of 
increasing the countercharge density and stabilizing the system against large variations of $\nu$, with $A$ 
and $B$ suitable constants. It is worth noticing that $A$ arises from the {\it short-range} part of the 
ion-ion interaction only because the Coulomb long-range part is exactly compensated by the electrons 
($n=\nu$). This short-range character and the absence of kinetic energy for the ions renders this term 
practically negligible (at least in comparison with the much larger inverse electron compressibility 
$\sim W$). Quite relevant (but hard to estimate from first principles) is the $B$ term acting when the 
ion density increases and stabilizing the ion system against a high-density collapse. 

Starting from Eq. (\ref{grandcanonical}), the average number of electrons per unit 
cell can be obtained as $n=-\partial_\mu\omega=2W^{-1}\left[\mu+\lambda-\varepsilon_0(\nu)\right]$, 
which is the result of the integration of Eq.\,(\ref{constraint}) and 
yields
$\mu=\frac{1}{2}Wn-\lambda+\varepsilon_0(\nu)$. The condition 
$$\partial_\lambda\omega=\nu-\frac{2}{W}\left[\mu+\lambda-\varepsilon_0(\nu)\right]=0$$ 
enforces the constraint $n=\nu$. Finally, imposing equilibrium with respect to $\nu$ gives
$$\partial_{\nu}\omega=\frac{2}{W}\left[\mu+\lambda-\varepsilon_0(\nu)\right]
\partial_{\nu}\varepsilon_0+\lambda+\partial_{\nu}\omega_0=0,$$ 
whence $\lambda=-\left(n\partial_{\nu}\varepsilon_0+\partial_{\nu}\omega_0\right)$. 

\begin{figure}[htbp]
\includegraphics[angle=-0,scale=0.3]{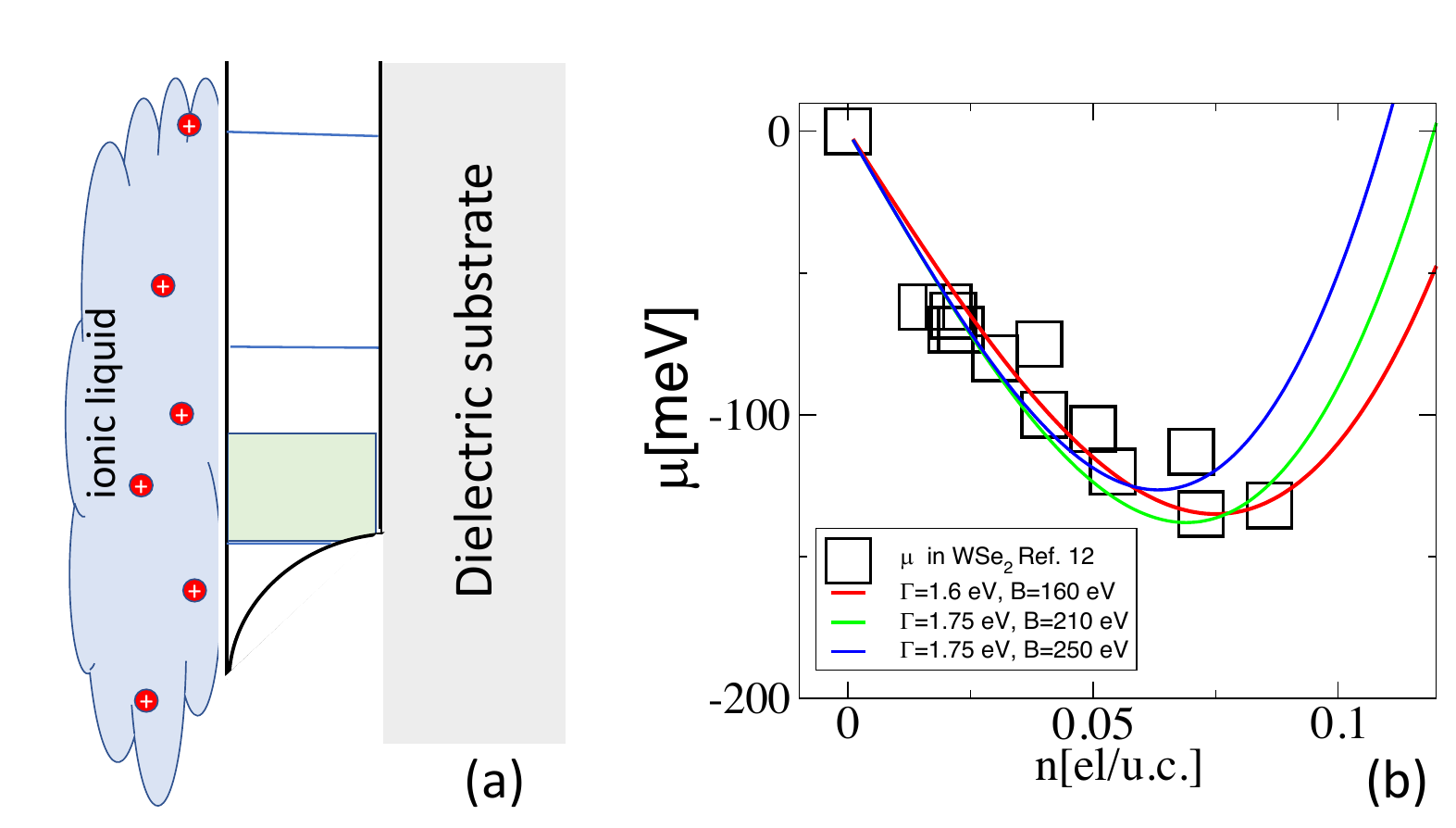}
\caption{(a) Schematic representation of the electronic potential well confining the 2D electron gas 
(green area) between the dielectric substrate (grey area) and the ionic liquid (light blue area). b) 
Chemical potential vs electron density in the 2D electron gas in the film. The parameters are $A=0$, 
$\Gamma=1.75$\,eV, $B=210$\,eV (green curve) and $B=250$\,eV (blue curve) and $\Gamma=1.6$\,eV, $B=160$\,eV 
(red curve). The square are data from Ref.\,\onlinecite{King-2015} obtained from ARPES experiments on surface 
doped WSe$_2$.
}
\label{EPS}
\end{figure}

Within an electrostatic continuous model it can easily be shown that the depth of the confining potential 
well increases linearly with $\nu$: the more abundant are the positive ions in the liquid gate side and 
the lower is the electrostatic confining energy trapping the electron gas at the surface. On the other hand, 
trapped charges may counteract this dependence when $\nu$ is small: A small density  of liquid ions
induces a corresponding small density of electrons, which, in this poorly screened situation,
are trapped as bound states around the positively charged liquid ions. In this situation the electronic level
$\varepsilon_0(\nu)$ is locked to the (random)  bound state levels.
To describe this low-density situation and its smooth evolution to the intermediate density situation, in which
$\varepsilon_0(\nu)$ linearly decreases because of the electrostatic interaction with the charged ions, 
 we phenomenologically write $\varepsilon_0(\nu)=-\Gamma\nu^2/(\nu+\nu_0)$, where $\nu_0$ is the threshold 
value above which a linear dependence is recovered and $\Gamma$ is a constant. The parameter $\Gamma$,
embodying the dependence of the bottom of the 2D electron band on the ion density, can be expressed in terms 
of the capacitance of the interface $\tilde C$, according to the relation $\Gamma=|e|/(2\tilde C a^2)$, where 
$e$ is the electron charge and $a$ is the lattice spacing of the 2D unit cell. Typical numbers are
$\tilde C \approx 10\,\mu$F\,cm$^{-2}$ and $a\approx 3\times10^{-8}$\,cm,\cite{wushi-2015} yielding 
$\Gamma \approx 10$\,eV. A more accurate (self-consistent) treatment of the potential well confining the 
electrons at the interface, which is beyond the scope of the present work, could provide a better estimate 
of the numerical prefactor relating $\Gamma$ to the typical interfacial potential scale $e/(\tilde C a^2)$. 
The typical electron bandwidth can be estimated as $W\approx 1$\,eV.\cite{wonseokyun-2012}
Putting together all the pieces, we can now write the electron chemical potential as 
\beq
\mu=\frac{W}{2}n+n(A+Bn^2)-\Gamma \frac{n^2(2n+3\nu_0)}{(n+\nu_0)^2}.
\label{mu}
\eeq
As it is readily seen, when $\nu_0=0$, and for $\frac{W}{2}+A-2\Gamma<0$, the inverse compressibility 
$\kappa^{-1}=\partial_n\mu$ is negative at small $n$, so that the system is unstable against electronic 
phase separation, as shown in Fig.\,\ref{EPS} for a typical parameter set that compares rather well with 
the negative compressibility observed in photoemission experiments in surface doped WSe$_2$. We point out 
that, if electronic phase separation is allowed to fully develop, the access to the negative compressibility 
region is forbidden, and the customary Maxwell construction must be used to determine the inhomogeneous 
composition of the phase-separated system. The fact that a negative electron compressibility is 
experimentally measured in surface doped WSe$_2$\cite{King-2015} may indicate that chemical dopants 
are not as mobile as ions in the liquid, thereby leaving the possibility that electronic phase separation 
is frustrated by residual long-range Coulomb forces, due to the lack of perfect compensation between 
electrons and countercharges. In this case, the short-range electron compressibility could stay negative, 
while the electron charge density is spatially distributed to find a compromise between the local tendency to 
electronic phase separation and the Coulombic cost of a charge-unbalanced density profile\cite{lorenzana}. A 
somewhat similar situation occurs in the model of Ref.\,\onlinecite{shklovski-2010}, which may be relevant 
at low electron density, or in badly metallic systems.

A finite $\nu_0$ would stabilise the compressibility at small $n$, while the term proportional to $B$ 
always stabilises the system against large variations of the density, yielding a finite density window 
where the system is unstable, and separates into two metallic phases, with higher and lower electron density. 
For $\nu_0=0$ the low-density phase is (band) insulating, while a finite $\nu_0$ may allow for a low-density
metallic phase to occur.

\begin{figure}[htbp]
\includegraphics[angle=-0,scale=0.3]{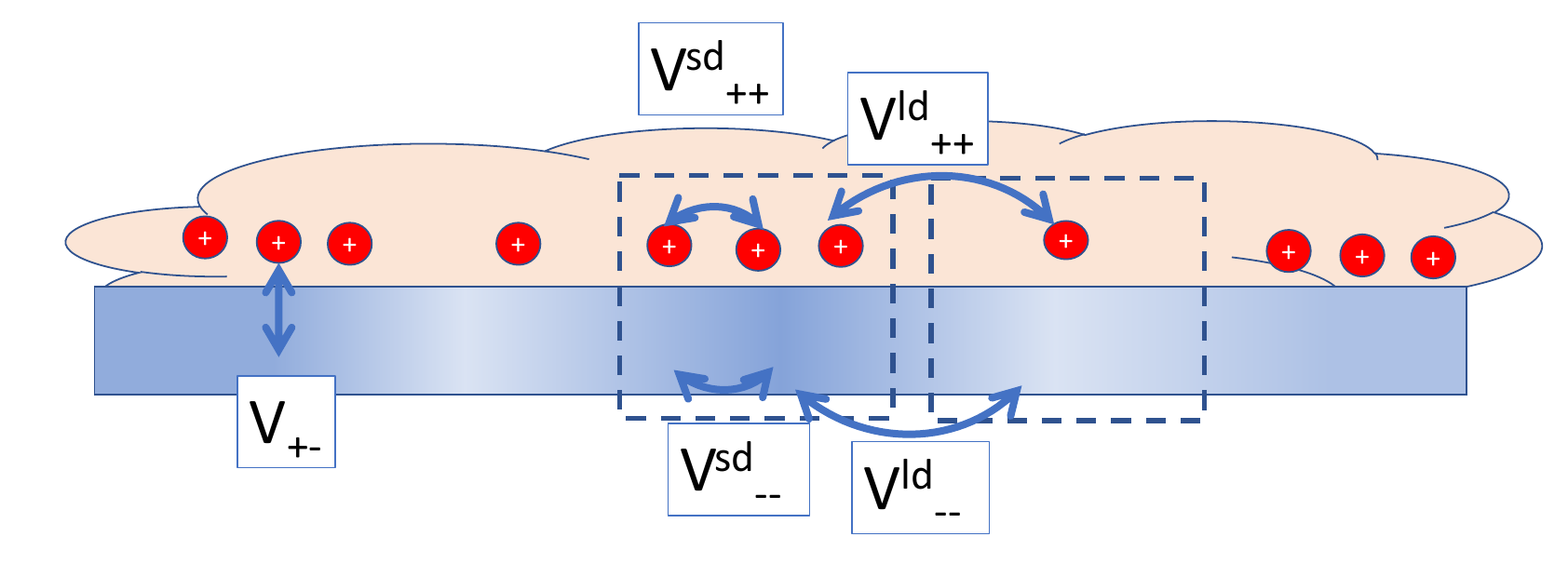}
\caption{Schematic representation of the inhomogeneous electronic distribution (blue and light blue 
regions) accompanied by the inhomogeneous ionic distribution (red circles) embedded in the liquid droplet. 
}
\label{fig4-scheme}
\end{figure}

{\section{ Discussion}}
Besides its formal implementation, the above model is a rather general representation of the physical 
situation schematically depicted in Fig.\,\ref{fig4-scheme}. The ion-ion repulsion is strongest at 
short-distances ({\it sd}), $V^{sd}_{++}$, preventing excessive ion accumulation in small volumes,
and within our phenomenological coarse-grained model this effect is modelled by the
rapid increase of the quartic term in the free energy [$B$ term in $\omega_0(\nu)$]. At 
large distances  ({\it ld}), despite the poor screening inside the liquid droplet, the underlying electrons balance the 
ionic charge and give rise to a small (dipolar-like) repulsion $V^{ld}_{++}$ (at larger scales the
inhomogeneous regions, represented by the dashed contours in Fig.\,\ref{fig4-scheme}, are essentially neutral). 
The same holds for the electron-electron repulsion, which is weak at large distances, $V^{ld}_{--}$, due 
to compensating ionic countercharges and may be larger at short distances, $V^{sd}_{--}$. For electrons there 
is the additional effect of mutual screening and high charge compressibility (large DOS of the 2D electron 
gas metallic state). In this scheme a large attractive contribution is present between the mobile electrons 
and the mobile (nearly frozen) ions at high (low) temperature. This electrostatic gain is such that the 
system may find it convenient to have a moderately higher ionic and electron densities to exploit this 
attraction. Of course, if the density fluctuation becomes large, the repulsive cost (i.e., the B term of 
the ion-ion free energy) stops the aggregation tendency. Therefore, besides the high 
screening inside the metallic layer, a negative electronic compressibility is favoured by the reciprocal 
charge compensation. Formally this is represented by the constraint $n=\nu$.

On a general basis we observe that an inhomogeneous state may result from rather generic attractive 
interactions induced by the interplay between the confined 2D electron gas and the countercharges coming 
from the gate, from chemical doping, or oxygen vacancies, as recently proposed for the oxide 
heterostructures.\cite{Bovenzi-2015,Scopigno-2016} The inversion-asymmetric electric field confining 
the 2D electron gas may also induce a strong Rashba spin-orbit coupling, which depends on the local 
electron density and may provide an additional source of effective electronic 
attraction.\cite{Caprara-2012,Bucheli-2014} These effective attractions can produce an electronic 
phase separation in the 2D electron gas as it is also supported by experimental evidences of a negative 
electronic compressibility in graphene-MoS$_2$ heterostructures,\cite{tutuc-2014} in WS$_2$,\cite{King-2015} 
in SrTiO$_3$ surface,\cite{Claessen-2016} and in LaAlO$_3$/SrTiO$_3$ interfaces.\cite{Luli-2011} Of course, 
specific features are present in each system that render the electronic phase separation specific. For 
instance, the size of the inhomogeneous regions depends on the frustrating effects of electron-electron 
and countercharge-countercharge Coulomb repulsions, which in turn depend on their mobility and on the 
screening in the various parts of the system. In the present case, we have considered a model where the 
charges of ionic liquid gating favor an extended electronic phase separation thanks to their high 
mobility (above their freezing temperature) that allows a large scale segregation of electrons and 
ions while keeping an overall quasi neutrality. We notice that our proposed mechanism may cooperate 
with electron inhomogeneity induced by the frozen ionic liquid locally detaching from the sample 
surface.\cite{Morpurgo-2015} Of course, when electrons are introduced by chemical doping the countercharges 
are much less mobile and a more evenly distribution of inhomogeneity (if any) is expected. This is why 
in this case the metal-to-superconductor transition is generically narrower and no tail is present 
in the $R(T)$ curves [see, e.g., Fig.\,\ref{EPS}(c,e) in Ref.\,\onlinecite{Bhoi-2017}]. Notice, however, 
that at low doping [Fig.\,\ref{EPS}(a) in Ref.\,\onlinecite{Bhoi-2017}] relative fluctuations in the 
distribution of the dopants become more important and again a broad transition is found. The situation 
may be even more intricate when competing phases like CDWs are present, as in the case of the domain wall 
formation recently discovered by scanning tunneling experiments\cite{Madhavan-2017} in Cu-intercalated 
1T-TiSe$_2$. Still it is worth noticing that also in this case our phenomenological RRN model can capture 
and describe the rather filamentary structure of the metallic domain walls responsible for incommensuration 
in the CDW and superconductivity in this system. We also notice that a suitable choice of the temperature 
dependence of the resistance in the normal metallic regions also allows a very good description of the onset 
of superconductivity in MoS$_2$ on amorphous substrates.\cite{Biscaras-2015}

Despite the variety of situations and of microscopic mechanisms giving rise to the inhomogeneity,
our phenomenological RRN model is quite general as long as the superconducting regions are large enough 
to sustain a local superconducting state. The physics of the system, at a mesoscopic level, is fully 
encoded by the spatial distribution of the cluster, its connectivity, and the parameters of the $T_c$ 
distribution. This is why upon reducing the average electron density, the model also allows to distinguish 
and describe the physically very different situations reported in Fig.\,\ref{resist} (a) and \ref{resist} 
(f,k), where superconductivity disappears because the puddles become sparse and less dense or because the 
local $T_{c}$ is degraded by the competition with CDWs. When the spatial distribution is filamentary, the 
low connectivity of the RRN accounts for the tailish shape of $R(T)$, with few non-superconducting 
puddles preventing the zero resistance state and $R(T)$ staying finite until the very last puddles become 
superconducting at low $T$. On the contrary, when the puddles are more evenly distributed, the connectivity 
is large, percolating paths are easier to find and $R(T)$ vanishes without the long tail. 

Our inhomogeneous scenario can be tested by critical current experiments: despite the complex structure 
of the superconducting cluster, in the proximity of the critical current, transport should be ruled by the 
weakest links coupling nearby puddles, so that the critical current and its temperature and magnetic 
field dependencies are expected to be well described by the behavior of a single (or a few) Josephson 
junction(s). This behavior has already been found in LaAlO$_3$/SrTiO$_3$ interfaces.\cite{guenevere-2016} 
In the same systems a loosely connected (filamentary) regime has also been identified in radio-frequency 
measurements of the dynamical conductivity to obtain the superfluid density.\cite{Singh-2017} We suggest 
that similar experiments in 2D crystalline superconducting systems would provide valuable information.

\vspace {0.5truecm}
\noindent {\bf Acknowledgments -- }
We acknowledge stimulating discussions with L. Benfatto, C. Castellani, and F. Mauri and 
valuable correspondence with Y. Saito and  B. Shklovskii.
This work has been supported by the Sapienza University with Project n. RM11715C642E8370

\vspace {1truecm}
\begin{appendix}

{\section{Construction of the random superconducting cluster}}
We give here some details on the construction of the RRN modelling the inhomogeneous 
2D crystalline superconducting systems discussed in Section II.

1) We start from a 2D array of resistors meeting at the nodes of a 2D $N\times N$ lattice 
(see Fig.\,\ref{scheme1B}(a)). The two extremal columns (1 and N) of this 
array represent the external leads at potential $V$ and $0$ respectively. Each bond $(i,j,\hat{x})$, along $x$, 
or $(i,j,\hat{y})$, along $y$, corresponds to a resistor. Each resistor represents a metallic or 
superconducting region connected to 6 neighboring regions as shown in the right-hand side of 
Fig.\,\ref{scheme1B}(b). A bond  $(i,j,\hat{e})$ is completely characterized by the current 
$I (i,j,\hat{e})$ flowing through it and by its resistance $R (i,j,\hat{e})$. The relation between 
the local resistance $R (i,j,\hat{e})$, the local critical temperature $T_c (i,j,\hat{e})$ and the 
temperature $T$ follows from the nature of the system intended to study. Here, the form of the relation
$R(i,j,\hat{e})=R[T;T_c(i,j,\hat{e})]$ depends on whether the effects of disorder should be introduced 
through a distribution of the critical temperatures and/or through a specific dependence of the local 
resistance on the (critical) temperature. Given that we will be mainly interested in the effects of 
the distribution of critical temperatures and of spatial correlations between the islands, we discard 
the latter possibility in order to avoid blurring these effects through complicated $T$ and $T_c$ 
dependencies. We thus consider the simplest case of a binary resistor:
\begin{eqnarray}
R(i,j,\hat{e}) &=& R_0 :T >T_c(i,j,\hat{e}),  \\
&& 0 :T <T_c(i,j,\hat{e})
\end{eqnarray}
This assumption is reasonable whenever the temperature range in which the resistance of a 
single resistor goes from its normal-state value $R_0$ to 0 is much smaller than the width of the 
$T_c$ distribution.\cite{CGBC} In addition, the normal-state resistivity of each superconducting bond is 
independent of its critical temperature. Taking $R_0$ finite means that an island is either metallic 
or superconducting, excluding the presence of insulating islands. As far as the effects of disorder 
on transport are concerned, the present approach is essentially classical; the superconducting regions 
are assumed to be large enough to have a stable fixed phase with fully established local coherence 
and negligible charging energy. As a consequence, neighbouring superconducting resistances immediately 
establish a mutual phase coherence as soon as both have become superconducting.

2) The second step consists in forming the regions of the 2D crystalline superconductor, which at low 
enough temperature become spatially correlated superconducting regions. It turns out that to fit the 
peculiar shape of the resistance curves of Fig.\,1, two geometric ingredients are needed. On the one hand, 
bulky regions must become superconducting to account for the smooth marked decrease of the resistance 
from the high-temperature side of the metal-to-superconductor transition. On the other hand, the tailish 
shape of $R(T)$ on the low-$T$ side of the transition can only be reproduced by weakly connected regions. 
To this purpose we proceed as follows [see also Sect. II and particularly Fig. \ref{scheme1B}]:
i) we generate a filamentary weakly connected fractal-like structure and then ii) we decorate this random 
nearly one-dimensional tree with larger 2D circular puddles. To generate the fractal-like structure we adopt a 
a simple growth process known as diffusion limited aggregation (DLA). Its construction is as follows: A particle 
is released at the left edge of a 2D lattice and let diffuse to the right. More precisely, the particle moves one 
bond to the right and then with equal probability one bond up or down. This sequence is iterated until the particle 
stops, as soon as it reaches the top, bottom or right edge, where it sticks. Then, other particles are launched one 
after another and halted either when reaching one of the three edges or a bond already occupied by one of the 
previously diffused particles. The cluster obtained in a $250 \times 250$ square lattice after diffusing 50000 
particles is defined by the bonds where the particles sticked. Due to a saturation at the left edge, the total 
number of superconducting bonds only amounts to about 25000. Once this large cluster is obtained (see 
Fig.\,\ref{scheme1B}(c)) we select a $100 \times 100$ sub-lattice as shown in the green part of 
Fig.\,\ref{scheme1B}(c). 
The restriction is made to yield a more physical case where the low-dimensional cluster covers the whole sample. 
So, henceforth we consider restricted lattices ranging from $100 \times 100$ to $200 \times 200$ sites, 
where only bonds belonging to the cluster are assigned a finite critical temperature $T_c$. The other 
bonds belonging to the rest of the 2D square-lattice grid are not reported in Fig.\,\ref{scheme1B}(c) for clarity and 
appear as the white background regions. In Fig.\,\ref{scheme1B}(d) they are instead reported in light yellow
and they form a resistive background with the typical 
resistivity $R_0$ at all temperatures. The filamentary skeleton of the system is then extended by bulkier 
2D patches of circular shape with a diameter ranging from 5 to 30 bonds leading to structures as shown in 
Fig.\,\ref{scheme1B}(d)

3) Once the spatially correlated structure of the superconducting subset of the 2D crystalline superconductor 
is formed, the random character of superconductivity in this subsystem remains to be implemented. The 
physical rationale is that, once intrinsic mechanisms induce density inhomogeneities, substantial fluctuations 
of the local superconducting properties can arise from a density-dependent local $T_c(1,n)$ or from fluctuations 
of the quenched disorder. The critical $T_c$ is the simplest physical property (and phenomenological parameter) 
that can reflect the randomly distributed superconducting properties. Thus we considered for the sake of 
definiteness the Gaussian distribution of $T_c$s reported in Eq.(\ref{gauss}). When the temperature is 
progressively decreased, the regions with the $T_c>T$ become superconducting enlarging the superconducting 
regions (always inside the chosen spatially correlated subset) according to the schematic view of 
Fig.\,\ref{scheme1B}(d).

Notice that the additional gate-potential axis is present in Fig.\,\ref{scheme1}: the increased average number 
of electrons is reflected in the increase of the total fraction $w$ of the superconducting regions 
[$P(T_c)$ being normalized to one].

4) With the above steps the resistances are set forming the desired geometrical structure of a mixture of 
filamentary and large puddles (superpuddles), equipped with a randomly chosen local critical temperature
a fixed voltage $V$ is applied at the two vertical edges of the lattice, while the horizontal edges have open 
boundary conditions. The physics of the RRN is governed by the usual equations of electrostatics,
$$
{\bf J}({\bf r}) = -\sigma ({\bf r}) {\bf \nabla} V ({\bf r}); \,\,\, {\bf \nabla} \cdot {\bf J}({\bf r}) = 0
$$
where ${\bf J }({\bf r})$ is the current density and $\sigma ({\bf r})$ is the conductivity at position ${\bf r}$. 
In the discrete case of a square lattice, these equations simplify to Ohm's and Kirchoff's laws
\begin{eqnarray}
R(i, j,\hat{x}) I(i, j,\hat{x}) &=& V (i +1, j)-V (i, j)  \nonumber \\
R(i,j,\hat{y}) I(i,j,\hat{y}) &= &V(i,j+1)-V(i,j)   \nonumber \\
\sum_{\langle \hat{e}\rangle}  I(i,j,\hat{e}) &=& 0,   \nonumber
\end{eqnarray}
where $V(i,j)$ is the electrostatic potential of site $(i,j)$ and the symbol $\langle \hat{e}\rangle$  
restricts the sum over all bonds surrounding a given node $(i,j)$. 
Implementing the above equations for each of the $N^2$ nodes and for each of the $2N^2 - 2N$ bonds, one obtains 
a set of $3N^2 -2N$ linear equations, which determines the $N^2$ voltages of the nodes and the $2N^2 - 2N$ 
currents of the bonds. In other words, one needs to solve a system of linear equations $\bf{Ab} = \bf{y}$ 
where the elements of the matrix ${\bf A}$ consist of either $\pm1$, $\pm R_0$ or $0$, the vector ${\bf b}$ 
contains the unknown potentials and currents, and the vector ${\bf y}$ contains the known terms which are 
either $V$ or 0. Once the system is solved, the total current $I$ flowing from one edge to the other can 
be calculated by summing the $N+1$ currents of any of the $N$ vertical links,
$$
I=\sum_{j=1}^N I(i,j,\hat{x}).
$$
Due to charge conservation, this sum is independent of $i$.
The ratio $V /I$ then determines the global resistance of the cluster at a given temperature. As the temperature 
is decreased, more and more bonds become superconducting and the global resistance decreases. If at a certain 
temperature $T_p$ a connected superconducting path joins the leftmost and the rightmost edge, the resistance of 
the network drops to zero, i.e., a percolative metal-to-superconductor phase transition occurs.
\end{appendix}



\begin{thebibliography}{99}

\bibitem{geim2013}A. K. Geim and I. V. Grigorieva, Nature 499, 419 (2013).

\bibitem{iwasa-2016} Yu Saito Tsutomu Nojima and Yoshihiro Iwasa, 
Nature Rev. {\bf 2}, 16094 (2017).

\bibitem{tinkham} M. Tinkham, {\it Introduction to Superconductivity}, (McGraw-Hill, New York, 1996).

\bibitem{biscaras-2013} J. Biscaras, N. Bergeal, S. Hurand, C. Feuillet-Palma, A. Rastogi, R. C. Budhani, M. Grilli, 
S. Caprara, and J. Lesueur, 
Nature Mater. {\bf 12}, 542 (2013).

\bibitem{bucheli-2013} D. Bucheli, S. Caprara, C. Castellani, and M. Grilli, 
New J. Phys. {\bf 15}, 023014 (2013).

\bibitem{caprara-2013} S. Caprara, J. Biscaras, N. Bergeal, D. Bucheli, S. Hurand, C.
Feuillet-Palma, A. Rastogi, R. C. Budhani, J. Lesueur, and M. Grilli, 
Phys. Rev. B {\bf 88}, 020504(R) (2013).

\bibitem{guenevere-2016} G. E. D. K. Prawiroatmodjo, F. Trier, D. V. Christensen, Y. Chen, N. Pryds, and T. S. Jespersen, 
Phys. Rev. B {\bf 93}, 184504 (2016).

\bibitem{shengchun-2016} Shengchun Shen, Ying Xing, Pengjie Wang, Haiwen Liu, Hailong Fu, Yangwei Zhang, Lin He, 
X. C. Xie, Xi Lin, Jiacai Nie, and Jian Wang, 
Phys. Rev. B {\bf 94}, 144517 (2016).

\bibitem{saito-2018} Yu Saito, Tsutomu Nojima, and Yoshihiro Iwasa, Nature Commun. {\bf 9}, 778 (2018).

\bibitem{Caprara-2012} S. Caprara, F. Peronaci, and M. Grilli, 
Phys. Rev. Lett. {\bf 109}, 196401 (2012).

\bibitem{Scopigno-2016} N. Scopigno, D. Bucheli, S. Caprara, J. Biscaras, N. Bergeal, J. Lesueur, and M. Grilli,
Phys. Rev. Lett. {\bf 116}, 026804 (2016).

\bibitem{tutuc-2014} Stefano Larentis, John R. Tolsma, Babak Fallahazad, David C. Dillen, Kyounghwan Kim, Allan H. 
MacDonald, and Emanuel Tutuc, 
Nano Letters {\bf 14}, 2039 (2014).
 
\bibitem{King-2015} J. M. Riley, W. Meevasana, L. Bawden, M. Asakawa, T. Takayama, T. Eknapakul, T. K. Kim, M. Hoesch, 
S.-K. Mo, H. Takagi, T. Sasagawa, M. S. Bahramy and P. D. C. King, 
Nature Nanotech. {\bf 10}, 1043 (2015).

\bibitem{Ye-2012} J. T. Ye, Y. J. Zhang, R. Akashi, M. S. Bahramy, R. Arita, and Y. Iwasa, 
Science {\bf 338}, 1193 (2012).
 
\bibitem{Saito-2015} Y. Saito, Y. Kasahara, J. Ye, Y. Iwasa, and T. Nojima, 
Science {\bf 350}, 409 (2015).
 
\bibitem{Li-2015}  L. J. Li, E. C. T. O'Farrell, K. P. Loh, G. Eda, B. \"Ozyilmaz, and A. H. C. Neto, 
Nature {\bf 529}, 185 (2015).

\bibitem{Bhoi-2017} D. Bhoi, S. Khim, W. Nam, B. S. Lee, Chanhee Kim, B.-G. Jeon, B. H. Min, S. Park, and 
Kee Hoon Kim, 
Sci. Rep. {\bf 6}, 24068 (2016).

\bibitem{CGBC} S. Caprara, M. Grilli, L. Benfatto, and C. Castellani, 
Phys. Rev. B {\bf 84}, 014514 (2011).

\bibitem{shklovski-2010} Brian Skinner and B. I. Shklovskii, 
Phys. Rev. B {\bf 82}, 155111 (2010).

\bibitem{Bovenzi-2015} N. Bovenzi, F. Finocchiaro, N. Scopigno, D. Bucheli, S. Caprara, G. Seibold, and M. Grilli,
Journal of Superconductivity and Novel Magnetism {\bf 28}, 1273 (2015).

\bibitem{Bucheli-2014} D. Bucheli, M. Grilli, F. Peronaci, G. Seibold, and S. Caprara, 
Phys. Rev. B {\bf 89},  195448 (2014).

\bibitem{Claessen-2016} Lenart Dudy, Michael Sing, Philipp Scheiderer, Jonathan D. Denlinger, Philipp Schütz, 
Judith Gabel, Mathias Buchwald, Christoph Schlueter, Tien-Lin Lee, and Ralph Claessen,
Adv. Mater. 2016, DOI: 10.1002/adma.201600046
 
\bibitem{Luli-2011} Lu Li, C. Richter, S. Paetel, T. Kopp, J. Mannhart, R. C. Ashoori,
Science {\bf 332}, 825 (2011).

\bibitem{Morpurgo-2015} Jo, S., Costanzo, D., Berger, H., and  Morpurgo, A. F., 
Nano Lett. {\bf 15}, 1197-1202 (2015).

\bibitem{Madhavan-2017} Shichao Yan, Davide Iaia, Emilia Morosan, Eduardo Fradkin, Peter Abbamonte, and Vidya Madhavan,
Phys. Rev. Lett. {\bf 118}, 106405 (2017).

\bibitem{Singh-2017}
G. Singh, A. Jouan, L. Benfatto, F. Couedo, P. Kumar, A. Dogra, R. Budhani, S. Caprara, M. Grilli, E. Lesne, 
A. Barth\'el\'emy, M. Bibes, C. Feuillet-Palma, J. Lesueur, N. Bergeal,
Nature Commun. {\bf 9}, 407 (2018).

\bibitem{Biscaras-2015} J. Biscaras, Z. Chen, A. Paradisi, and A. Shukla, 
Nature Commun. {\bf 6}, 8826 (2015).

\bibitem{zhu2004}L. Zhu, P. J. Hirschfeld, and D. J. Scalapino, Phys. Rev. B {\bf 70}, 214503 (2004).

\bibitem{landauer} R. Landauer, {\it Electrical Transport and Optical Properties of Inhomogeneous Media}, 1978 edited  by J. C. Garland and
D. B. Tanner (New York: AIP). 

\bibitem{kirkpatrick} S. Kirkpatrick, Rev. Mod. Phys. {\bf 45}, 574 (1973)

\bibitem{joshua-2012} Arjun Joshua, S. Pecker, J. Ruhman, E. Altman, and S. Ilani, Nature Commun. 3, 1129 (2012).

\bibitem{biscaras-2012}J. Biscaras, N. Bergeal, S. Hurand, C. Grossetete, A. Rastogi, R. C. Budhani, 
D. LeBoeuf, C. Proust, and J. Lesueur, Phys. Rev. Lett. 108, 247004 (2012).

\bibitem{finkelstein-1987} A. M. Finkelstein, Sov. Phys. JETP Lett. 45, 46 (1987)

\bibitem{wushi-2015} Wu Shi, Jianting Ye, Yijin Zhang, Ryuji Suzuki, Masaro Yoshida, Jun Miyazaki, Naoko Inoue, Yu Saito, 
and Yoshihiro Iwasa, 
Sci. Rep. {\bf  5}, 12534 (2015).

\bibitem{wonseokyun-2012} W. S. Yun, S. W. Han, S. C. Hong, I. G. Kim, and J. D. Lee, 
Phys. Rev. B {\bf 85}, 033305 (2012).

\bibitem{lorenzana} J. Lorenzana, C. Castellani, C. Di Castro, 
Phys. Rev. B {\bf 64}, 235127 (2001); J. Lorenzana, C. Castellani, and C. Di Castro, 
Phys. Rev. B {\bf 64}, 235128 (2001);
C. Ortix, J. Lorenzana, and C. Di Castro, 
Phys. Rev. B {\bf 73}, 245117 (2006).

\end{thebibliography}
\end{document}